# Two methods for Jamming Identification in UAVs Networks using New Synthetic Dataset

Joseanne Viana †‡, Hamed Farkhari∗†, Luis Miguel Campos ∗, Pedro Sebastião †‡,
Francisco Cercas †‡, Luis Bernardo §‡, Rui Dinis§‡,

†ISCTE – Instituto Universitário de Lisboa, Av. das Forças Armadas, 1649-026 Lisbon, Portugal

∗PDMFC, Rua Fradesso da Silveira, n. 4, Piso 1B, 1300-609, Lisboa, Portugal

‡IT – Instituto de Telecomunicações, Av. Rovisco Pais, 1, Torre Norte, Piso 10, 1049-001 Lisboa, Portugal

§FCT – Universidade Nova de Lisboa, Monte da Caparica, 2829-516 Caparica, Portugal;

Emails : joseanne_cristina_viana@iscte-iul.pt, Hamed_Farkhari@iscte-iul.pt, luis.campos@pdmfc.com,
pedro.sebastiao@iscte-iul.pt, francisco.cercas@iscte-iul.pt, lflb@fct.unl.pt, rdinis@fct.unl.pt

*Abstract*—Unmanned aerial vehicle (UAV) systems are vulnerable to jamming from self-interested users who utilize radio devices for their benefits during UAV transmissions. The vulnerability occurs due to the open nature of air-to-ground (A2G) wireless communication networks, which may enable network-wide attacks. This paper presents two strategies to identify Jammers in UAV networks. The first strategy is based on time series approaches for anomaly detection where the signal available in resource blocks are decomposed statistically to find trend, seasonality, and residues, while the second is based on newly designed deep networks. The joined technique is suitable for UAVs because the statistical model does not require heavy computation processing but is limited in generalizing possible attack's identification. On the other hand, the deep network can classify attacks accurately but requires more resources. The simulation considers the location and power of the jamming attacks and the UAV position related to the base station. The statistical method technique made it feasible to identify 84.38% of attacks when the attacker was at 30 m from the UAV. Furthermore, the Deep network's accuracy was approximately 99.99 % for jamming powers greater than two and jammer distances less than 200 meters.

*Index Terms*—Cybersecurity, Convolutional Neural Networks (CNNs), Deep Learning, Jamming Detection, Jamming Identification, UAV, Unmanned Aerial Vehicles, 4G, 5G;

I. INTRODUCTION

When it comes to the 5G communication system, the deployment of unmanned aerial vehicles (UAVs) is a gamechanger. They allow faster and more flexible network services at higher data rates in the sky because they have complete control over their movement and high probability of establishing robust line-of-sight (LoS) communication links[1], [2] and [3]. Yet, UAV transmission is vulnerable to attacks and interference because of the open nature of air-to-ground (A2G) wireless communication links and the A2G channel connections that may present opportunity for attacks in the network. As a result, it is critical and vital to identify threats and protect UAV communications[4] . In wireless communication, security is ensured through encryption and encoding techniques commonly employed to prevent unwanted access and intentional interference. Despite this, maintaining encryption systems requires a lot of effort and resources [5]. As a result, encryption in UAVs' transmission may not be feasible. Therefore attacks identification mechanisms become fundamental in UAV networks. Commonly used jamming detection algorithms such as packet delivery ratio and received signal strength-based detection with a miss detection rate are presented by [6].

Statistical models have lately been recognised as a feasible methods for monitoring network activity in wireless communications and detecting suspicious attacks via the use of wireless channel properties rather than encryption keys. In [7], the authors propose a jamming detection method using a Naive Bayes classifier trained on a limited sample of data considering only noise effects in the transmission in wireless scenarios. Cheng et al [8] describes a Bayesian method for jamming detection. In [9], the authors offer a jamming detection strategy for GNSS-based train localization that makes use of singular value decomposition (SVD). Lu et al. [10] developed a technique for detecting jamming in power networks that is both efficient and resilient. Most of the studies made no account for the impacts of the channel in the computation. Considering machine learning and Deep Networks, Youness et al.[11] analyze the signal properties that may be used to detect jamming signals, create a dataset based on these parameters and used the random forest method, the support vector machine algorithm, and neural network algorithm to classify the features extracted by the jamming signal. Li et al. [12] also identify jamming samples using signal-extracted features but the author adds another way to detect attacks that utilizes 2D samples and pretrained networks (i.e AlexNet, VGG-16, ResNet-50). Although, with pre-trained networks, we may utilize transfer learning to adapt the network to a new dataset without having to design from scratch, certain pre-trained networks are enormous and need significant computational processing in order to categorize information which may be unsuitable for UAVs. While embedded statistical models and deep network techniques at the cloud or edge can monitor and analyze channel degradation caused by jamming attacks, there is a lack of publicly available research on attack detection in UAV communications. Rather than identifying attacks, most research in this field focuses on prevention, namely anti-jamming measures and non-traditional ways to avoid jamming.

We aim to demonstrate that it is feasible to identify attacks in the receiver block of the UAV by combining a Seasonal Trend Decomposition (STL) time series analyser with a unique deep network architecture that has much fewer layers than the well known pretrained networks and does not rely on transfer learning techniques.

The remainder of the paper is structured as follows: First, the detailed system model is presented in Section II. It describes the dataset used, the statistical approach, and the suggested deep network architecture. The results related to the performance assessment of both approaches are summarized in Section III. Finally, section IV concludes this paper.

Notations: Scalar variables are denoted by lower-case letters (a, b,...), vectors are denoted by boldface lower-case letters (a, c,...), and matrices are denoted by boldface capitals (A, B,...). Lower case letters denote time-domain variables, whereas upper case letters indicate frequency-domain variables.

*A. Contributions and Motivation*

With regard to jamming detection and associated challenges utilizing deep networks and statistical methods, there is a lack of public research and accessible data. Taking this into consideration, the following highlights some of the contributions made by this paper:

- A general case study model that takes into account the jamming power and distances between the jamming attacker and base station related to the UAV that uses the average received signal power in the resource block, Signal-To-Noise-Ratio (SNR), average-noise, averagetransmitted-power, path-loss and shadowing.
- A statistical model for jamming detection using data from the UAV reception resource block.
- A simpler Convolutional Neural Network-Long Short Term Memory (CNN-LSTM) architecture for jamming detection.
- Simulation results on the both presented techniques.
- A comparison of two jamming detection techniques' performance in terms of accuracy over attacker distance and power.

Additionally, we offered a visual representation of the confusion matrix for both the training and test data sets. We devised a strategy that increases performance while using the fewest CNN layers.

## II. SYSTEM MODEL

We analyse a UAV jamming scenario in which a communication link exists between the base station and the UAV, referred to as an air-to-ground (A2G) connection and there are jamming attackers in unknown locations on the ground or in the air that can deliberately jam the signal received by the authenticated UAV. Although we use the Single Carrier (SC) transmission scheme, the jamming detection algorithms are applicable to any transmission technique. We investigate the reception power in the authenticated UAV using two distinct approaches, one of which relies on time series statistical models and the other on deep networks. Each component is explained as follows:

*A. dataset*

The data set simulates the received signal in the UAV resource block considering small fading and shadowing effects in the transmission channel. The frequency domain channel $H_{(k,d)}$ is represented as in the equation 1,

$$H_{(k,d)} = \frac{\sum_{i=1}^{N_{rays}} \alpha_i(\tau) + \exp(-j2f\pi\tau)}{\sqrt{(PLS)}} \quad (1)$$

For $H_{(k,d)}$, $f$ is the frequency band, $\alpha$ is the attenuation of the multipath ray, and $\tau$ is the propagation delay. We assume the Rician model to describe the multipath rays. The small fading or path loss is estimated by the UAV and base station locations $p_{id,t} = [x_{uav,t}, y_{uav,t}, z_{uav,t}]T$, $[x_b, y_b, z_b]T$ (in meters) and the 3D Euclidean distance $||p_{bs}-p_{uav}||^2$ equation respectively. The reference point is at $d = 10m$ and $S$ or shadowing is a random variable modeled as $S|_{db} \sim N(0, \sigma^2)$. The received power $Y_{(k,d)}$ is calculated using equation 2,

$$Y_{(k,d)} = H_{(k,d)}X_k + N, \quad (2)$$

$X_k$ is the frequency domain representation of the transmitted signal $x_k^R$, and $N \sim N(\mu, \sigma_k^2)$ is the noise in the channel, while $k$ is an available frequency in the bandwidth.

The jamming signal takes into account the same properties of the reliable signal considering path loss. In the experiment, the jamming focuses on $F_k \ll F$ frequencies inside the bandwidth $B$ available for transmission with power gain $(P/P_J)I$, where I is the percentage of the slot occupied by the jammer. The jammer is more powerful than the signal in the majority of the dataset samples (i.e. $P_J > P_S$). Equation 3 shows how the total noise is affected by the jamming power received, where $N_k$ is the noise without interference.

$$E[|N_{k,Tot}|^2] = \frac{F}{F_{k,J}} \frac{P_j}{P_S} E[|N_k|^2]. \quad (3)$$

The dataset contains 483,540 transmission blocks or samples $Sa$, with N steps each $\{Sa_k; k = 0,1,…,N − 1\}$ where N is the FFT size in the frequency domain. The received signal is then classified according to the following categories: Good-Normal, Bad-Normal, Good-Jamming, and Bad-Jamming. "Normal" states for non jamming signal and the $SNR = 20$ and $SNR = 1$ distinguish Good and Bad channels, respectively. Additionally, we vary the jammer and the base station position as well as the jammer power in the experiment. Figure 1 depicts two jamming sample available in the dataset. In the top, Jamming in a good channel. In the dow, jamming in a bad channel. (the dataset contains only the Received power related to the four classes mentioned previously).

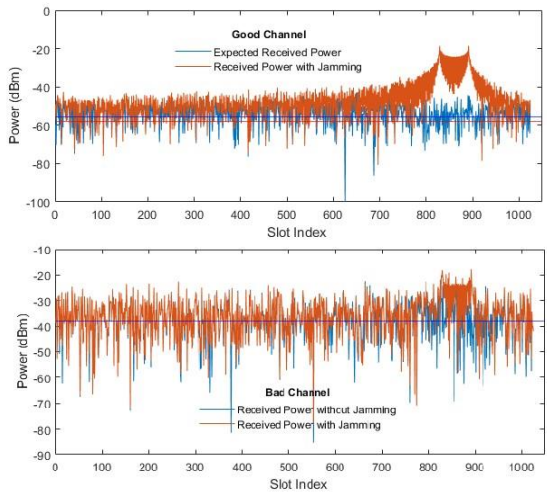

Fig. 1: Received signal with jamming in Good and Bad Channels.

The table I compares the mean received power difference at the UAV when the signal is jammed and when the signal is not jammed with the same power as the authorized signal at a distance of 30m from the UAV. The base station is about the same distance from the UAV. It is obvious that the jammer modifies the mean power in the resource block, depending on the jammer's power and position relative to the base station and the UAV.

TABLE I: Mean Difference Between Channels with and Without Jamming

| Mean Power difference (dBm) | No Jamming | Jamming |
| --- | --- | --- |
| Good Channel | 0 | 4.189 |
| Bad Channel | 0 | 0.592 |

### B. Statistical methods

The statistical model chosen was STL [13]. It takes into account the decomposition of the signal into trend, seasonal and residuals. The figures 2 and 3 illustrate a representative sample of both jammed and unjammed decomposed signals. The top chart in both figures shows a combination of three samples in a sequence from the dataset. In figure 2, the jamming power of attacker is five times greater than the signal received from the base station. The jamming location is $30m$ away from the UAV, while the UAV placement is $90m$ away from the base station. In figure 3, there is no jamming attacker and the base station location is identical to that in figure 2.

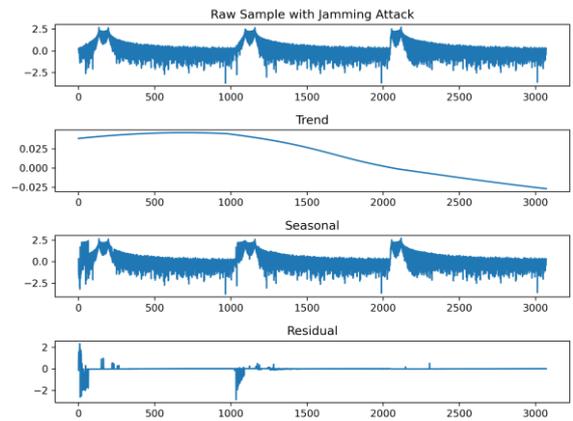

Fig. 2: STL decomposition of a normalized sample in the present of jamming attacker.

STL is an acronym for "Seasonal and Trend decomposition based on locally weighted regression (Loess)" and is a technique for decomposing time series into trend, seasonal, and residual components. After removing the current trend estimate, the seasonal component of the cyclic sub-series is calculated using Loess interpolation (seasonal smoothing). The predicted seasonal component is then smoothed using another Loess interpolation (lowpass smoothing). Finally, the deseasonalized series is smoothed once more (trend smoothing) in order to provide an estimate of the trend component. This procedure is repeated numerous times in order to improve the component estimates' accuracy[13]. Each sample contains elements that degrade the communication in the experiment, such as noise and slow fading components. Additionally, the channel and the noise effect are independent of each other. On the other hand, the jamming effect will manifest itself in multiple samples in the case of a jamming attacker. We exploit the jamming attack's periodicity to identify it using STL. We assumed the jamming attack is applied in a certain narrow channel bandwidth up to (8% to 10% ). After the third resource block reception, we combine three samples in a sequence, apply the STL decomposition, and then reconstruct the signal again using equation 4; we then calculate the amount of error for each sample as in equation 5. If there is a pattern in the sample, the number of errors is smaller; consequently, the sample is classified as a jamming signal. On the other hand, in a normal situation without a jamming effect, there is no specific repeated pattern, and we see more errors in the STL decomposition reconstructed signal. At the end, we use root mean square error (RMSE) for binary classification of the presence of jamming effect and the absence of it in normal situation in equation 6. Finally, we applied Support Vector Machine (SVM), Logistic Regression and Random Forest algorithms for split the features in the classes. The difference between the dataset described in the previous section and the one used in the STL is the concatenation of three resource blocks in one sample. While elaborating the experiment, we noticed that the *period* is a fundamental parameter to be chosen in order to get good results from STL.

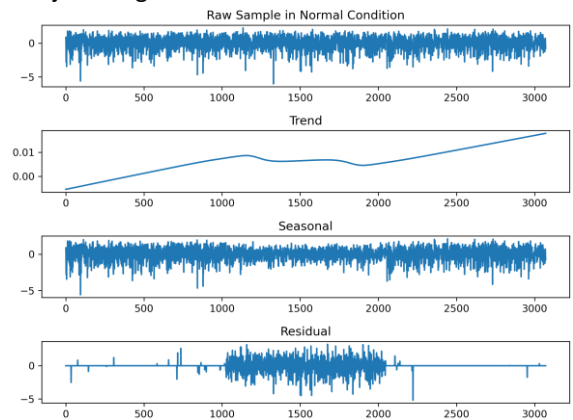

Fig. 3: STL decomposition of a normalized sample without jamming signal.

$$S_a = T + S + R \quad (4)$$

$$Error = S_a - S_r \quad (5)$$

$$RMSE = \sqrt{\frac{1}{N}\sum_{i}^{N} Error_i^2} \quad (6)$$

In the equations 4, 5 and 6, $S_a$ and $S_r$ are the original and reconstructed samples, and $T$, $S$, and $R$ are stands for trend, seasonal, and residual components, respectively. $N$ is length of sample and we used *RMSE* as a feature for binary classification.

*C. Convolutional Neural Networks*

In the experiment using convolutional neural network (CNN) joined with Long short-term memory (LSTM), we developed an architecture capable of achieving 99 % accuracy with a small number of CNN layers, number of filters and kernel sizes in the convolutional layers. Our CNN architecture uses three convolutional layers, one LSTM, one drop-out, a fully connected, and the output layer for classification as illustrated in Figure 4. A common layer in CNN is the maxpooling but when it is used in certain topologies, it might result in the loss of critical information and according to Geofrey Hinton: "pooling is a mistake". For this reason, we replaced the pooling layer with strides in our convolutional layers. Authors in [14] and [15] reported that using very small weight decay (L2 regularization) values such as $5 \times 10^{-4}$, and $4 \times 10^{-5}$ in convolutional layers are critical for performance purposes and it should be precisely chosen. After grid search, we found the optimal L2 regularization and used it for all CNN layers in the experiment.

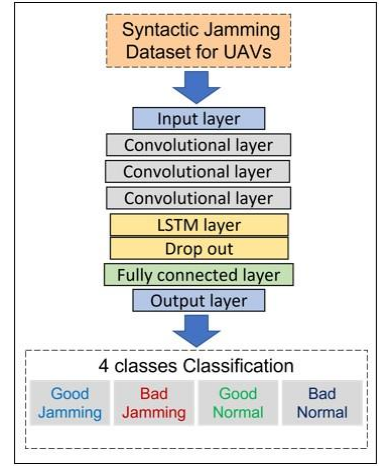

Fig. 4: Proposed CNN-LSTM Architecture.

For deep networks, we used a single sample dataset. The data set was initially partitioned into 70% for training and 30% for testing. In the first phase, we divided the training section into two sub-sections: train and validation for the grid search algorithm used to determine the deep network's hyperparameters. The hyperparameters are as follows: the number of CNN filters, the kernel sizes, the strides, the batch size, the learning rate, the number of regularization terms, and the drop out percentage. In the second phase, we employed a 5-fold cross validation procedure during the training phase to ensure accuracy and avoid overfitting.

III. EXPERIMENTAL RESULTS

The results of the suggested algorithms are detailed below. First, we look at the statistical data for jamming detection. Then we'll look into deep networks for classification, loss and accuracy in train and test. The CNN model were trained and tested in a system with a Nvidia RTX 3090 GPU. The jamming attacker signal power ratio ranges between one and twenty. The distances between the UAV and the base station, and UAV and jamming attacker varies between 10 and 350 meters. Shadowing variance was adjusted to 4 [16].

*A. Statistical model*

The figure 5 depicts the RMSE between original signal and reconstructed signal after STL decomposition in the BoxPlot. The diagram shows that both distributions can be splited using binary classification. We merged three resource block in a sequence which each of them is the size of $N = 1024$ in length, and we used $N$ as a period parameter for STL decomposition algorithm. After calculating the reconstructed signal, we used RMSE as a feature for detecting the jamming attack over distance and power. Based on the jamming attacker power ratio and the distances of UAV from base station and attacker, the overall accuracy of this method for all scenarios using the three different classifiers was about 70%, as is shown in Figure 6. In some cases, depending on the jammer and base station location relative to the UAV, the accuracy can be increased to up to 84.38% by employing an SVM classifier that outperforms two other classifiers.

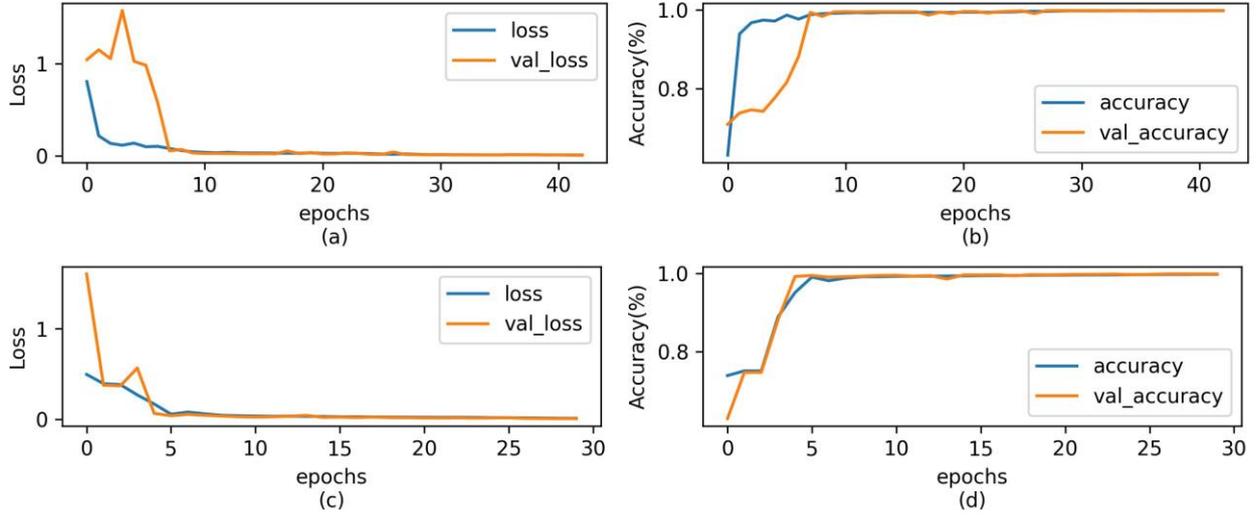

Fig. 8: Convergence of deep network during training with 4 steps. a) loss of model 1. b) *accuracy* of model 1. c) loss of

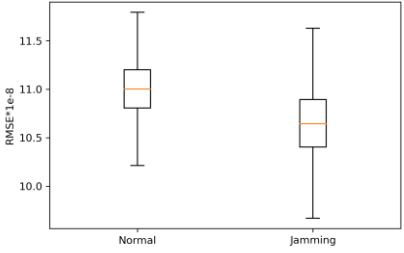

Fig. 5: Boxplot of RMSE between original signal and reconstructed signal for two classes.

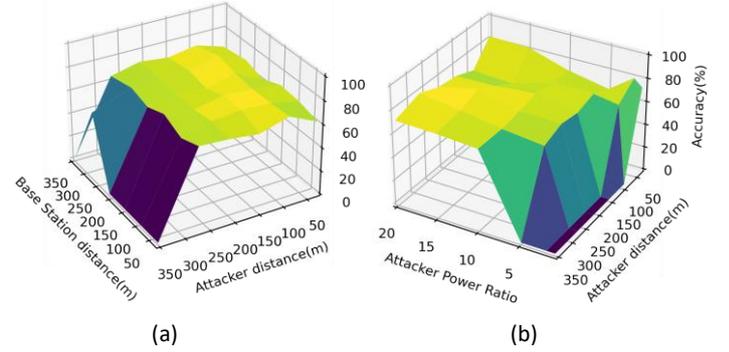

(a)                     (b)

Fig. 7: (a) Accuracy for $P_j = 5P_s$. (b) Accuracy for $Bs_d = 350m$.

model 2. d) *accuracy* of model 2.

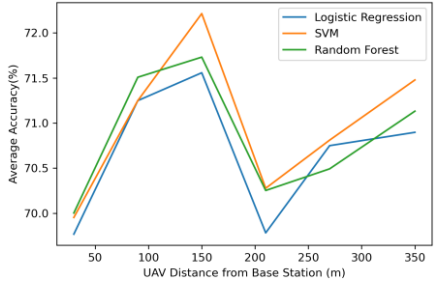

Fig. 6: Overall statistical average accuracy for different attacker powers and distances per fixed UAV distance from base station.

The figures 7 (a) and (b) ilustrate the *accuracies* of the STL model at various distances between the UAV and the attacker and at various attacker power ratios using the SVM classifier. In (a), the accuracy falls with increasing jammer distance, i.e. when the jammer is 350m away, the accuracy in the statistical model is reduced When the jamming power $P_j$ decreases as specified in (b), we see that it is difficult for the algorithm to differentiate low-power jammers due to the prominent channel effects such as fading, path loss, and shadowing.

Due to the statistical model's low computational requirements, it may readily be implemented to UAVs for user packet transmission and Command and Control (C2) links.

### B. Convolutional Neural Networks

In the CNN simulation, the slot size was set to N=1024. The only parameter used for the jamming detection was the signal received in the resource block. During the CNN performance configuration phase, it became clear that adding layers was preferable to increasing the number of filters in each layer. Additionally, the regularization component of the CNN and fully connected layer is crucial for achieving improved performance since its parameters must be appropriately adjusted. The LSTM layer was added to take advantage of the sequence memory characteristics in order to increase robustness. Also, the filter numbers and the kernel sizes implies in the overall trainable parameters. We achieved the same performance by employing two fully connected layers of 50 nodes each rather than a single layer of 100 nodes. These adjustments result in a decrease in the total number of trainable parameters of around 100k to 53k. Table II present the hyperparameters of our deep network.

TABLE II: Deep Network Configuration Parameters.

| Deep network Parameters | Value |
|---|---|
| base learning rate | $3.16 \times 10^{-3}$ |
| base batch size | 32 |
| conv-1 filters, kernel size, strides | 4, 8, 4 |
| conv-2 filters, kernel size, strides | 4, 4, 2 |
| conv-3 filters, kernel size, strides | 4, 3, 1 |
| LSTM | 100 |
| drop-out | 0.4 |
| dense | 100 |
| softmax | 4 |

We used L2 regularization terms equal to 1e-6 and 1e-5 in Convolutional layers and fully connected layer for both kernels and biases, respectively. An initial batch size of 32 TABLE III: Confusion matrix of 4 classes classification for test data by CNN-LSTM network.

| | Good Normal | Bad Normal | Good Jamming | Bad Jamming |
|---|---|---|---|---|
| Good-Normal | 36281 | 0 | 0 | 0 |
| Bad-Normal | 0 | 36219 | 0 | 62 |
| Good-Jamming | 0 | 0 | 36281 | 0 |
| Bad-Jamming | 0 | 385 | 0 | 35896 |

was used for training deep network, then we found the base learning rate of 3.16e-3 by using grid search algorithm. After that, we increased the learning rate and batch size to 0.2, and 2048, respectively. The new batch size and learning rate increased the GPU (RTX 3090 with 24GB Ram) use from 30% to 92% almost processing limit capacity; consequently, the batch size was limited to 2048. Following that, we trained our deep network in different steps of validation accuracy. In each training step, if the performance declined compared with the previous step, the training process was immediately stopped, the previous model weights was loaded, and the training process at that step was repeated with new lower learning rate and batch sizes. These steps were 80, 90, 95, and 99.99% of validation accuracy, and when the training process progressed through each step, we saved the model and continued the training process. We overcame the overshooting effect in deep network training and considerably shortened the overall training time for five models in 5-fold cross validation by employing this strategy. In Figure 8 as an example, the convergence of two models from 5 models in cross validation illustrated. It shows with maximum 40 epochs 99.99% *accuracy* achieved in each model.

Table III shows the confusion matrix of test set by the CNNLSTM algorithm. We were able to obtain corrected classification for all samples with good channels using the suggested deep network, and there was minimal misclassification in the case of bad channels. Specifically, we had 62 misclassifications in the absence of jamming and 385 when jamming was present TABLE IV: The result of test set for 4-Classes classification by proposed deep network.

| | precision | recall | f1-score | support |
|---|---|---|---|---|
| Good-Normal | 1.00 | 1.00 | 1.00 | 36281 |
| Bad-Normal | 0.99 | 1.00 | 0.99 | 36281 |
| Good-Jamming | 1.00 | 1.00 | 1.00 | 36281 |
| Bad-Jamming | 1.00 | 0.99 | 0.99 | 36281 |

which represents less than 1% of the total samples analysed. More specific details of precision and f-score parameters for the test provided in Table IV.

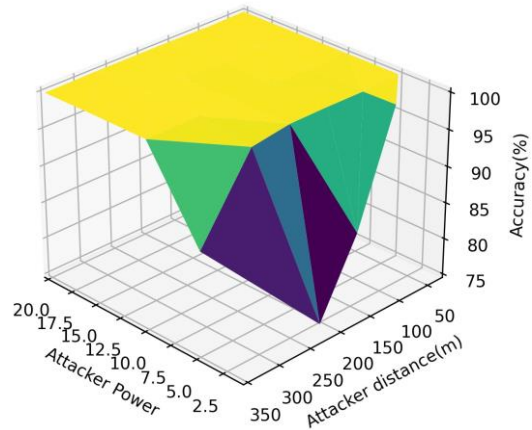

Fig. 9: CNN Accuracy for base station distance 30m.

To reduce the total number of trainable parameters, one of the following setup techniques may be used: Three CNN layers with eight filters each, followed by a LSTM and two fully connected layers with fifty nodes each. Alternatively, Three CNN layers with four filters each followed by a LSTM and only one fully connected layer with one hundred nodes. We discovered that the second one may converge more quickly and with fewer epochs.

Figure 9 depicts the CNN model's performance related to the *accuracy* over a range of distances between the UAV and the attacker and over a range of attacker power ratios. CNN may struggle to identify low-power jammers depending on the channel situation, as we saw in the statistical model, but in all other circumstances, CNN obtained 99.99% percent correct classification.

IV. CONCLUSION

This article offered a solution composed by two techniques for identifying jamming attacks in UAVs networks. The first one is based on time series approaches for detecting patterns by STL decomposition method, while the second is based on the convolutional neural networks. The signal analysed by both approaches rely on resource block received by the UAV. Using the time series analysis, it was possible to identify 84.38% of the attacks when the *SINR* of Jamming signal is high and UAV is closer to attacker than the base station. While using the deep networks the accuracy was 99.9% of the jamming cases and the false alarm occurred less than 1% of the cases. The combined method is appropriate for UAVs since the statistical model does not need extensive computer resources, but it is restricted in its ability to classify all conceivable attacks, while the deep network can classify all attacks but requires additional resources.

ACKNOWLEDGMENT

This research received funding from the European Union's Horizon 2020 research and innovation programme under the Marie Sklodowska-Curie Project Number 813391"